\documentclass[aip,rsi,reprint,graphicx]{revtex4-1}

\usepackage{graphicx}
\usepackage{dcolumn}
\usepackage{amsmath}
\usepackage{amssymb}
\usepackage{bm}
\usepackage[bbgreekl]{mathbbol}
\usepackage{color} 
\usepackage{braket}
\usepackage{gensymb}
\usepackage[dvipsnames]{xcolor}
\usepackage{lpic}
\usepackage[percent]{overpic}
\usepackage{stackengine}
\usepackage{physics}
\usepackage[para,online,flushleft]{threeparttable}
\usepackage{siunitx}
\usepackage{textcomp}
\usepackage{lipsum}
\usepackage{hyperref}
\hypersetup{
    colorlinks=true,
    linkcolor=blue,
    filecolor=magenta,
    citecolor=blue,
    urlcolor=black,
}
\usepackage[mathscr]{eucal}
\usepackage{ulem} 
\usepackage{amsfonts}

\usepackage[colorinlistoftodos]{todonotes}
\usepackage{mathtools}
\usepackage{amsmath}
\usepackage{empheq}
\usepackage{esvect}
\DeclareSymbolFontAlphabet{\mathbbm}{bbold}
\DeclareSymbolFontAlphabet{\mathbb}{AMSb}
\usepackage{multirow}
\usepackage{rotating}
\usepackage{booktabs}

\begin{document}

\preprint{APS/123-QED}

\title{A high performance active noise control system for magnetic fields}

\author{Tadas Pyragius}
\email{t.pyragius@gmail.com}

\author{Kasper Jensen}
\email{kasper.jensen@nottingham.ac.uk}

\affiliation{$^{1}$School of Physics \& Astronomy, University of Nottingham, University Park, Nottingham NG7 2RD, United Kingdom}


\begin{abstract}
We present a system for active noise control (ANC) of environmental magnetic fields based on a Filtered-x Least Mean Squares (FxLMS) algorithm. The system consists of a sensor that detects the ambient field noise and an error sensor that measures the signal of interest contaminated with the noise. These signals are fed to an adaptive algorithm that constructs a physical anti-noise signal cancelling the local magnetic field noise. The proposed system achieves a maximum of 35~dB root-mean-square (RMS) noise suppression in the DC-1~kHz band and 50~dB and 40~dB amplitude suppression of 50~Hz and 150~Hz AC line noise respectively for all three axial directions of the magnetic vector field.
\end{abstract}
\maketitle

Magnetically low noise environments are important across many metrologically relevant areas ranging from medical imaging of biomagnetic fields from the heart and brain to non-destructive evaluation of car batteries and scanning electron microscopy to name just a few \cite{fenici,Hu,sem}. Currently, optically pumped magnetometers (OPMs) are state of the art magnetic field sensors and are a promising alternative to conventional SQUID and fluxgate magnetometers in both shielded and unshielded conditions \cite{budker,mags}. A wider adoption of quantum magnetometers for ultra-low field precision measurements has been limited to magnetically shielded environments due to large external magnetic field noise, making such setups expensive. Conventionally, partial magnetic field noise cancellation is achieved via a combination of proportional-integral-derivative (PID) controllers which are sometimes combined with a feed-forward system \cite{active_PID,simple_PID0,simple_PID,mag_canc,mag_can2}. Whilst these systems are effective in field noise compensation and control they require extensive hands on tuning which is heavily dependent on the implemented hardware constraints. As a result, the tuning parameters cannot be easily transferred from one system to another. Furthermore, the performance of the noise suppression can be additionally limited by hardware and processing delays and system imperfections which often cannot be compensated for. In cases where the focus is on unwanted periodic environmental noise signals e.g. 50~Hz AC line noise and its corresponding harmonics, adaptive approaches have been implemented \cite{anc2,vigilante}. However, as before, some of these methods require extensive manual tuning of the filter coefficients as well as limiting the compensation to fixed AC noise frequency signals. Finally, it is often the case that the noise environment as well as the transfer function of the system are not known in advance or cannot be adequately modelled rendering these methodologies unable to cope. \\
\begin{figure}[t!]
\begin{overpic}[width=\columnwidth]{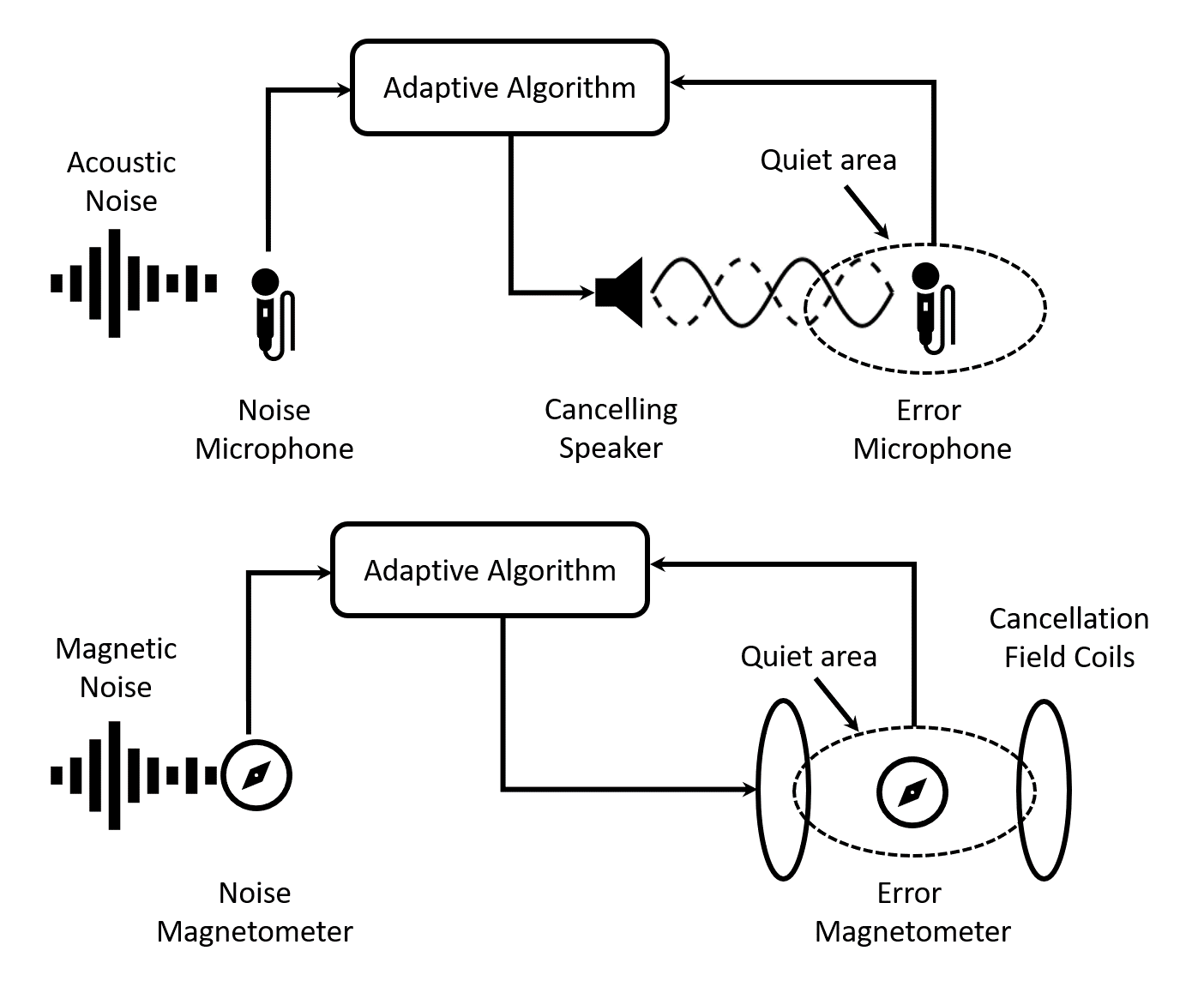}
\put(0,78){(a)}
\put(0,38){(b)}
\end{overpic}
\caption{(a) Basic implementation of active noise control in the acoustic domain. Here a noise reference microphone listens to unwanted environmental noise and feeds that information to the adaptive algorithm. The adaptive algorithm then calculates the anti-noise signal and outputs it through a noise cancelling speaker. The error microphone measures the noise + anti-noise signal and feeds it back to the adaptive algorithm for adjustment if necessary. (b) Analogous implementation of active noise control in the magnetic field domain. Here the microphones are replace by field sensitive magnetometers and the compensation is achieved by magnetic field coils.}
\label{fig:fig_sim}
\end{figure}

In this work, we demonstrate how these technical issues can be overcome using an active magnetic field compensation method based on adaptive filtering. Active noise control (ANC) is achieved by introducing a canceling “anti-noise” signal through a secondary source. The secondary source is driven by an electronic system which utilises a specific signal processing algorithm (such as an adaptive algorithm) for the particular cancellation scheme, see Fig.~\ref{fig:fig_sim}. This technique is widely exploited in noise cancelling headphone technology, vibration control, as well as exhaust ducts in ventilation and cooling systems \cite{headphones,interferometer_anc,anc1,duct}. Whilst adaptive filtering techniques have been demonstrated in unwanted noise cancellation of electric and magnetic fields in the context of electrocardiography (ECG) and magnetocardiography (MCG), the noise cancellation was performed on the acquired data \cite{ecg,mcg_opm_anc}. This process is known as adaptive noise cancellation \cite{widrow}. In contrast, active noise control (ANC), generates a physical anti-noise signal \cite{anc1,anc2}. This results in a number of advantages over adaptive noise cancellation techniques. First, the the noise signals in the environment are typically orders of magnitude larger than the signals of interest (e.g. magnetocardiography signals in $~$pT range compared to the 50~Hz line noise in $~$nT). This results in the requirement for a larger dynamic range which reduces the signal resolution of the signals of interest due to the limited number of bits in the analog-to-digital conversion (ADC). The ANC system cancels the physical noise signals which enables one to reduce the analog input range for the same number of bits, thus increasing the signal resolution and reducing the noise floor. In the context of OPMs, and in particular, spin-exchange relaxation-free (SERF) OPMs one of the key feature that determines the sensitivity of an OPM is its intrinsic linewidth \cite{serf}. When the environmental noise (e.g. 50 Hz noise) is larger than the OPM linewidth, this broadens the linewidth and reduces the sensitivity. ANC reduces the ambient magnetic field noise which in turn reduces the OPM linewidth increasing its sensitivity thus enabling its operation in otherwise environmentally noisy conditions. \\

This paper is organised as follows; Section I briefly introduces the formalism of signal discretization in the context of digital signal processing which is used throughout this work to compute the secondary path effects and calculate the corresponding anti-noise signal using adaptive algorithms. Section II and III walks through the motivation and process of secondary path modelling and its application to anti-noise generation via the filtered-x Least Mean Squares (FxLMS) algorithm. Section IV contains the results and discussion of the active noise control system for magnetic field noise cancellation, its scope, limitations and potential improvements. The last section sets out the conclusion.

\section{The z-transform for discrete time signals}
A discrete time signal can be represented by a weighted sum of delayed impulses such that
\begin{equation}
    x(n)=\sum_{k=0}^{\infty} x(k)\delta(n-k), \;\; \forall\; n, k \in \mathbb{Z}^+.
\end{equation}
To express the above more conveniently and remove the impulse response part, we can introduce a transform kernel which maps the impulse response with a delay to
\begin{equation}
    \delta(n-k)\rightarrow z^{-k}
\end{equation}
which allows us to transform the discrete time signal $x(n)$ in terms of variable $z$
\begin{equation}
    X(z)=\sum_{k=0}^{\infty}x(k)z^{-k},\;\; \forall\; k \in \mathbb{Z}^+.
\end{equation}
This kernel transformation enables easy signal manipulation using operations such as convolution. In discrete system control a signal input with a system's transfer function producing an output is a convolution of the input signal $x(n)$ and the system transfer function $t(n)$ which in a discrete z-transform is nothing but a product of the two functions, i.e.
\begin{equation}
    (x\ast y)[n]=\sum_{k=0}^{\infty} x(k)t(n-k)\rightarrow X(z)T(z).
\end{equation}
Due to the complexity of adaptive filters they cannot be trivially implemented on analog electronic platforms. As a result, adaptive filter technology is usually deployed on digital platforms such as application specific integrated circuits (ASICs) acting as digital signal processing (DSP) chips or field-programmable gate arrays (FPGAs). These digital platforms deal with discrete signals and thus require the mathematical formalism of discrete time signals.

\begin{figure}[t!]
\begin{overpic}[width=0.5\textwidth]{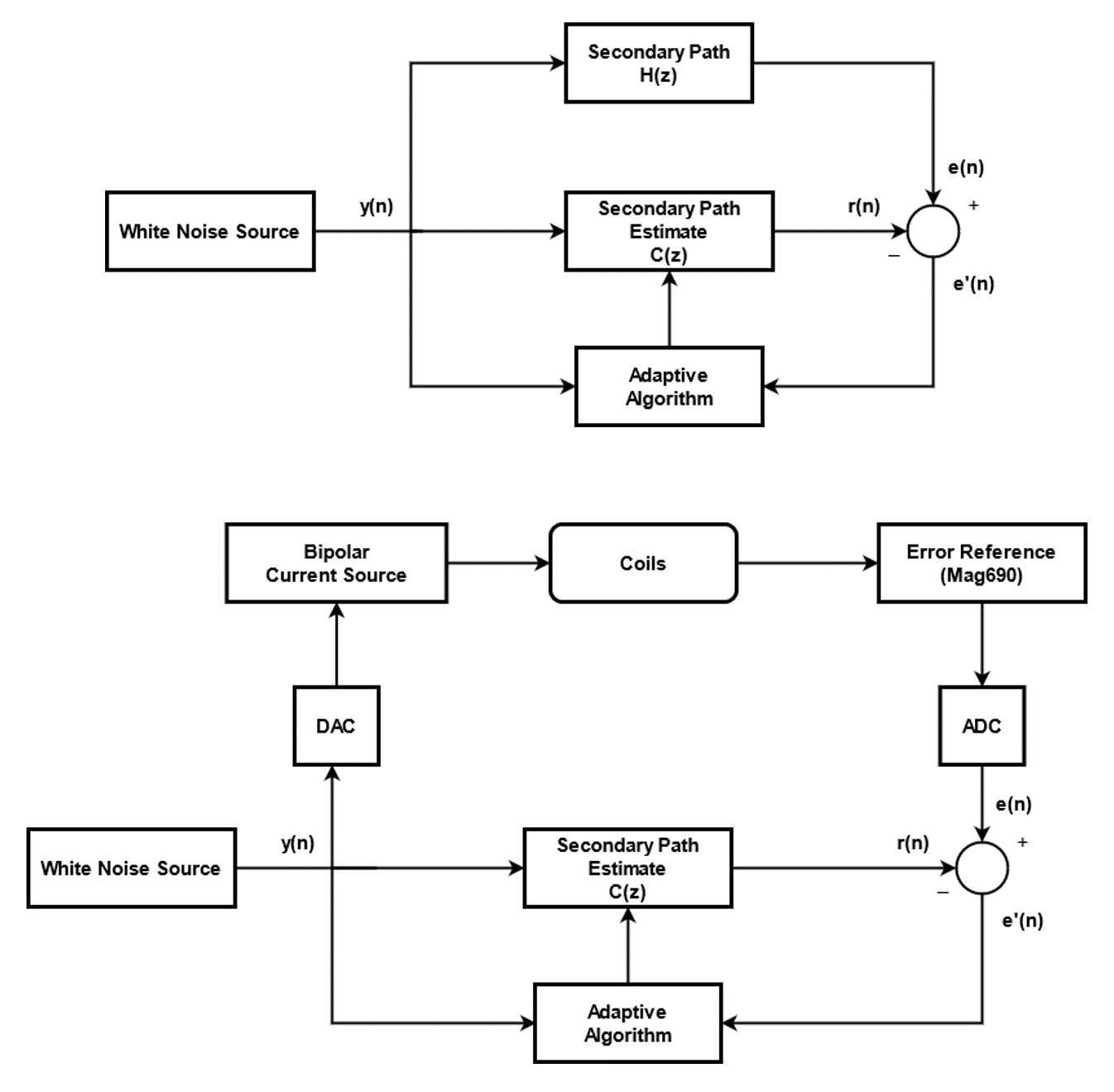}
\put(0,93){(a)} \put(0,49){(b)}
\end{overpic}
\caption{(a) Theoretical structure of the secondary path $H(z)$. (b) Experimental structure of the secondary path in the context of magnetic field generation and measurement. Here the white noise signal is fed to the digital-to-analog converter (DAC) driving a bipolar current source (BCS) with uniform field coils. The same white noise signal is fed to the adaptive algorithm as a reference. An error reference magnetometer measures the generated white noise field which is then used to calculate the error signal and adjust the secondary path coefficients. Here the secondary path $H(z)$ consists of a DAC, BCS, coils, magnetometer and an ADC with an anti-aliasing filter (not shown).}
\label{fig:sp_model}
\end{figure}

\begin{figure*}[t!]
\begin{overpic}[width=0.8\textwidth]{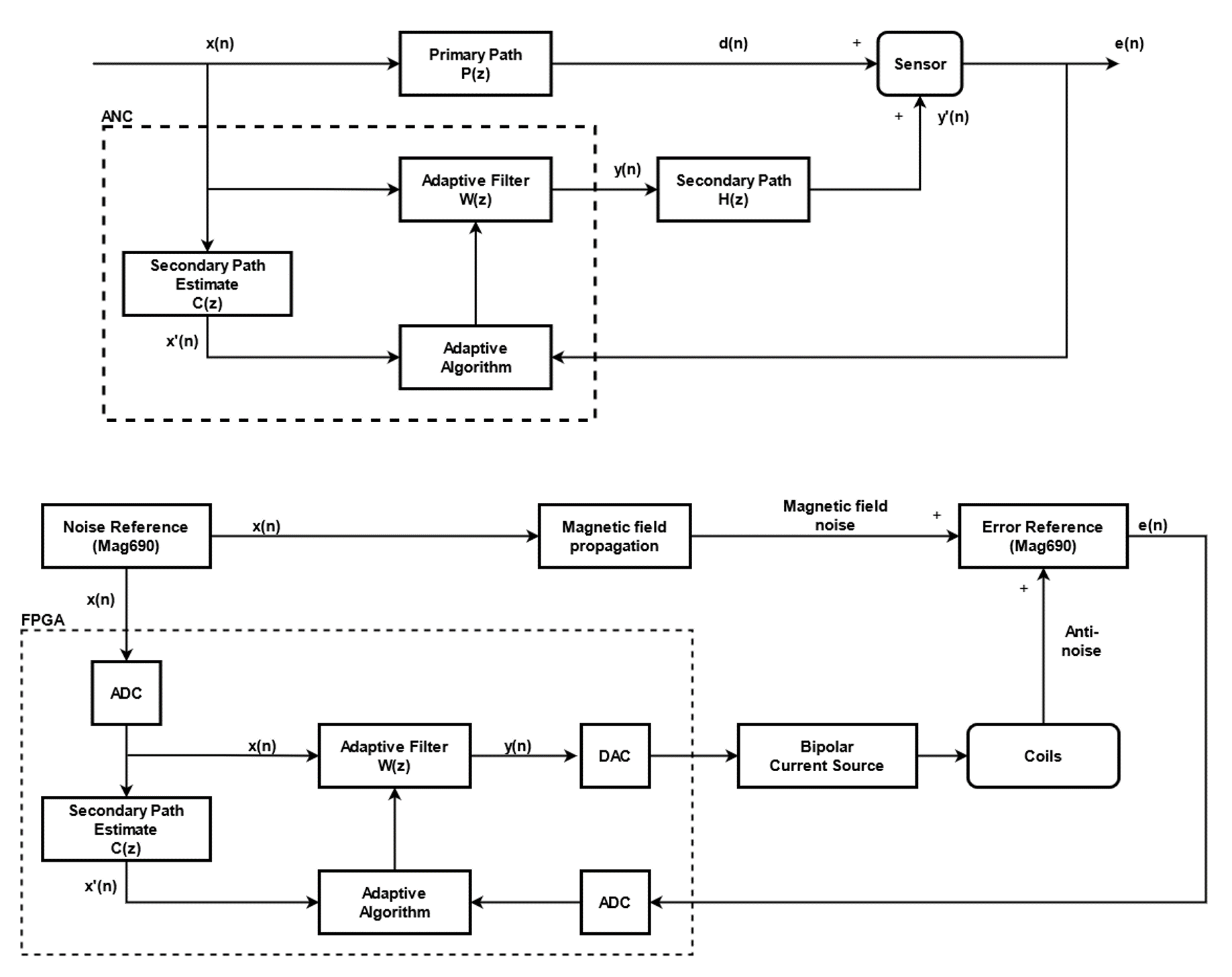}
\put(0,78){(a)} \put(0,40){(b)}
\end{overpic}
\caption{(a) Theoretical structure of the primary and secondary paths. The ANC box contains the algorithmic elements of the adaptive filter that generates an anti-noise signal. Here we assume that the primary path $P(z)$ transfer function does not have an appreciable effect on the signal detected at the sensor. (b) Experimental structure of the primary and secondary paths in the context of magnetic field generation and measurement. The ANC algorithm is contained within the FPGA chip.}
\label{fig:anc_model}
\end{figure*}

\section{Secondary Path Modelling}
For a broadband ANC system, the first stage of implementation consists of modelling the secondary path, see Fig.~\ref{fig:sp_model}~(a). The secondary path takes into account of the hardware effects of the output for a given input signal. The generated signal is detected by a sensor with a transfer function which modifies the measured signal. This signal is then propagated through the system which further modifies the signal. We want to compensate for these secondary path effects. This requires one to estimate the approximate secondary path response i.e. we want to find filter coefficient transfer function which can effectively approximate the transfer function of the secondary path, i.e., $C(z)\approx H(z)$, which corresponds to discrete transfer functions of the estimated and actual secondary paths respectively. To do this, we generate a white noise signal, $y(n)$ which satisfies
\begin{align}
    &\mathrm{E}[y(n)]=0, \\
    &\mathrm{Var}[y(n)]=\sigma^2,\\
    &\mathrm{E}[y(n)\ast y(n-k)]=0,\;\; \forall\; k \in \mathbb{Z}^+,
\end{align}
where E is the expectation value, Var is the variance with a value $\sigma^2$ and the last term is the auto-correlation between the two time delayed white noise signals. The white noise signal is fed to a bipolar current source which drives a pair of Helmholtz coils. Simultaneously, we use the generated $y(n)$ white noise as a reference for the secondary path estimate filter $C(z)$ and the least-mean-squares (LMS) adaptive algorithm. The generated white noise magnetic field is sensed by a fluxgate magnetometer (Bartington Mag690) and after an anti-aliasing filter is fed into an analog-to-digital converter. The sensed magnetic field corresponds to our secondary path response $e(n)$. For the adaptive model, the computed response $r(n)$ is given by
\begin{equation}
    r(n)=\sum_{i=0}^{M-1}c_i(n)y(n-i),
\end{equation}
where $n$ is the discrete time increment index, $c_i(n)$ is the $i^\textrm{th}$ filter coefficient of the discrete transfer function of the estimated secondary path $C(z)$ (equivalent to an adaptive filter) and $M$ is the total number of taps in the adaptive filter. An adaptive filter is a finite-impulse response (FIR) filter. The number of filter taps relate to the frequency resolution of the filter by $f_\mathrm{res}=f_s/M$, where $f_s$ is the sampling rate. The coefficients of the adaptive FIR filter $c_i(n)$ are the quantized values of the impulse response of the frequency transfer function \cite{taps}. The computed response $r(n)$ of the adaptive filter is then compared with the measured response of the system $e(n)$ which allows us to compute the error $e'(n)$
\begin{equation}
    e'(n)=e(n)-r(n).
\end{equation}
The measured error can then be used to update the filter coefficients of the secondary path estimate $C(z)$
\begin{equation}
    c_i(n+1)=c_i(n)+\mu_\textrm{sp}e'(n)y(n-i),
\end{equation}
where $\mu_\textrm{sp}$ is the step size of the update during secondary path modelling and must satisfy \cite{anc2}
\begin{equation}
    0<\mu_\textrm{sp}<\frac{1}{MP_y},
\end{equation}
where $P_y$ is the power of the signal $y(n)$. With the updated values of the coefficients $c_i(n)$ and a new secondary path estimate $C(z)$ the procedure is repeated iteratively to compute the new error value and update the corresponding coefficients. After some time, the iterative procedure is stopped. The final computed secondary path coefficients $c(i)$ and the estimated secondary path are used in the next stage of active noise control using the FxLMS algorithm.

\section{Filtered least mean squares (F\lowercase{x}LMS) algorithm}
With the estimated effect of the secondary path the next stage involves actively compensating for the noise. A reference noise signal $x(n)$ is propagated through the estimated secondary path $C(z)$ which yields the filtered version of the reference noise signal
\begin{equation}
    x'(n)=\sum_{i=0}^{M-1}c(i)x(n-i).\label{eq:1}
\end{equation}
We then proceed to compute the anti-noise signal
\begin{equation}
    y(n)=\sum_{i=0}^{M-1}w_i(n)x(n-i),\label{eq:2}
\end{equation}
where $w_i(n)$ is the $i^\textrm{th}$ coefficient of the adaptive filter $W(z)$. The adaptive coefficients are then updated according to
\begin{equation}
    w_i(n+1)=w_i(n)-\mu_\textrm{anc}e(n)x'(n-i),\label{eq:3}
\end{equation}
where $e(n)$ is the error reference signal and $\mu_\textrm{anc}$ is the ANC filter update step size, where typically $\mu_\textrm{sp}\neq \mu_\textrm{anc}$. The procedure is repeated. See Fig.~\ref{fig:anc_model} for schematic details.

\section{Experimental Implementation}
\subsection{Outline of the ANC procedure}
Applications of active noise control in noise cancelling systems generally deal with AC coupled signals (e.g. acoustic or vibration signals). This becomes problematic when a signal acquires a DC component which is the case in ANC applications of magnetic field cancellation due to an Earth's field component. A DC component affects the secondary path analysis by biasing the filter coefficients which render the cancellation system ineffective. As a result, in order to obtain a good estimate for the secondary path, $C(z)$, one must cancel the DC component of the ambient magnetic field and only then proceed with the online secondary path estimation. This step is achieved by a PID controller which pre-stabilises the magnetic field around zero. The output voltage sent to the bipolar current source that approximately cancels the Earth's field is then used as a DC control during the secondary path analysis combined with the white noise reference signal for impulse response modelling with the PID switched off. This ensures that the generated magnetic field signals are centered around zero without any DC offset. The secondary path modelling is run sequentially for each magnetic field vector axis individually for a given duration $T_\textrm{sp}$ and a step size $\mu_\textrm{sp}$. The estimated secondary path coefficients are then used to construct a secondary path response filter which enables the FxLMS algorithm to calculate the anti-noise signal in active noise control. The algorithm is deployed on NI's sbRIO-9627 FPGA and controlled using LabVIEW with NI's proprietary adaptive filter toolkit \cite{labview}.  

\subsection{Secondary Path Estimation}
\begin{figure}[t!]
\centering
\includegraphics[width=\columnwidth]{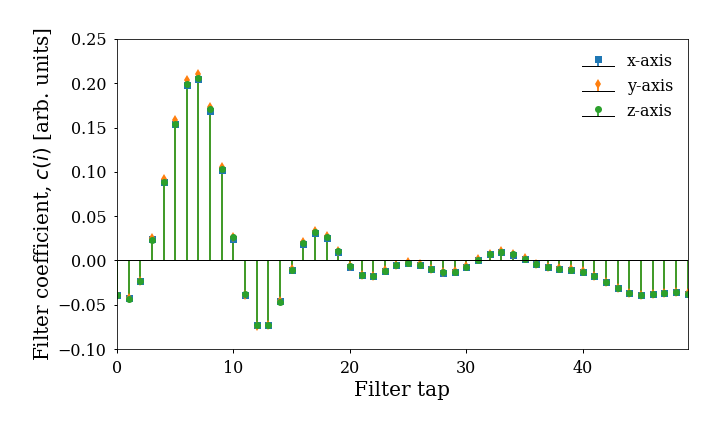} 
\caption{Experimentally obtained secondary path estimates for all three axial directions. Here we have the finite impulse response of a filter as a function of filter taps. The number of filter taps relate to the frequency resolution of the filter by $f_\mathrm{res}=f_s/M$, where $f_s$ is the sampling rate. We observe that the impulse response is shifted because the filter only operates on available samples with some additional delay. For an ideal low-pass filter, the filter coefficients follow a $sinc$ function and the delay is given by $(M-1)/2f_s$. The secondary path consists of a DAC, bipolar current source (BCS), fluxgate (Mag690), anti-aliasing low-pass filter (LP) and ADC. In this case the specification for each hardware channel is the same giving rise to identical secondary path coefficients.  }
\label{fig:sp_exp}
\end{figure}

The secondary path estimation follows an experimental setup depicted in Fig.~\ref{fig:sp_model}~(b). A band limited white noise signal is generated within the FPGA architecture and is used to drive a bipolar current source with a pair of Helmholtz coils producing a uniform white noise magnetic field around zero. The generated magnetic field is sensed by the error reference fluxgate magnetometer which is then fed in to the LMS algorithm on the FPGA to compute the error and adjust the filter coefficients which attempt to minimise the error. The filter coefficients allow one to construct an estimate of the impulse response of the secondary path, see Fig.~\ref{fig:sp_exp} for results. The profile of the secondary path impulse response depends on a number of various physical properties of the system such as anti-aliasing filters used for the ADCs, transfer functions of the coils, bipolar current source etc., as well as the relative gains of the input and output signals. In our experimental setup the 3-axis system has identical hardware and software implementation including the anti-aliasing filters and the coils (near identical resistance, inductance, number of turns). Consequently, the estimated secondary paths using the same experimental parameters are to a good approximation equivalent. 

\subsection{Active Noise Control using FxLMS algorithm}
\begin{figure}[t!]
\begin{overpic}[width=\columnwidth]{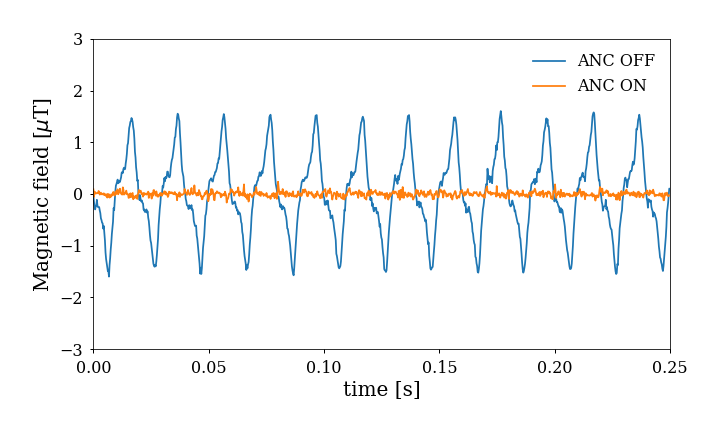}
\put(0,55){(a)}
\end{overpic}
\begin{overpic}[width=\columnwidth]{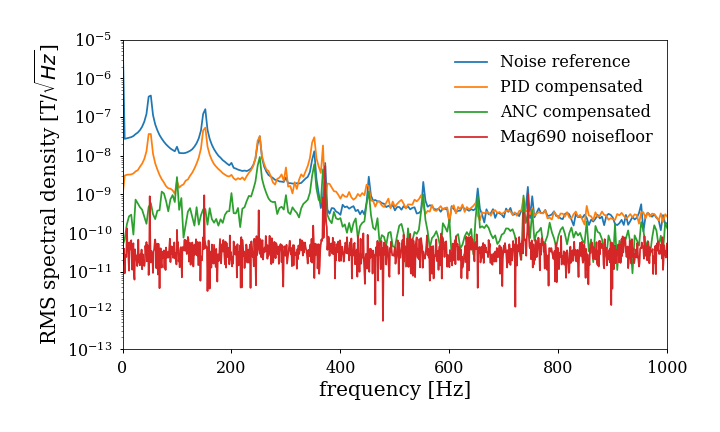}
\put(0,55){(b)}
\end{overpic}
\begin{overpic}[width=\columnwidth]{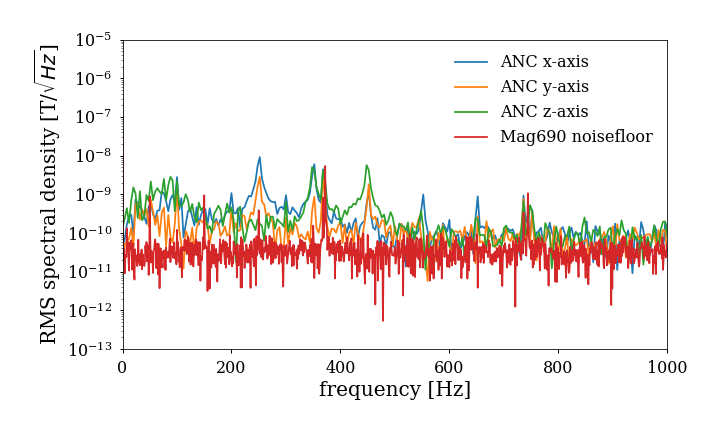}
\put(0,55){(c)}
\end{overpic}
\caption{(a) Time trace of the error magnetometer signal before and after applying active noise control. The time trace for the ANC OFF case has been offset to subtract the DC Earth's field for visualisation. (b) Root mean square amplitude spectral density of the magnetic field signals. The noise reference fluxgate which monitors the ambient field environment. ANC compensated is the magnetic field measured by the error magnetometer which monitors the cancelled magnetic field noise. The PID compensated noise is the error magnetometer measuring the noise performance using conventional proportional-integral-differential controller. (c) ANC performance RMS amplitude spectral density for all three axial directions run simultaneously. The Mag690 noisefloor is obtained inside a 4-layer $\mu$-metal shield in an open loop configuration.}
\label{fig:anc_results}
\end{figure}

The experimental layout of the FxLMS algorithm for active noise control is shown in Fig.~\ref{fig:anc_model} (b). Here the reference and error fluxgate magnetometers read the ambient and compensated magnetic fields which are fed to the FxLMS algorithm. The FxLMS algorithm compensates for the secondary path, adjusts the filter coefficients and computes an anti-noise signal $y(n)$ (see eqns.~\ref{eq:1}-\ref{eq:3}), which drives the bipolar current source controlling a pair of Helmholtz coils to produce a uniform anti-noise magnetic field. The active noise control system is initialised sequentially for each magnetic field vector axis. Once the compensation reaches a steady state limit the ANC system is disabled and the compensation is initialised for the next axis until a steady state is reached there and so on. Once each axial direction has converged with optimum filter coefficients, the ANC system is then switched on simultaneously for all three axial directions. In this configuration the ANC algorithm is capable of compensating for AC as well as DC fields without requiring additional Earth's field nulling. The initial DC field nulling is only required during the secondary path estimation. Due to cross-talk between each axis direction, some field leakage occurs and the FxLMS algorithm has to re-adjust the optimum filter coefficients when operated in 3-axis mode. Fig.~\ref{fig:anc_results} (a) shows the time trace comparison between the ambient field noise with and without the ANC engaged with Figs.~\ref{fig:anc_results} (b-c) detailing the frequency band performance of ANC for each vector direction. As can be seen from the figure the ANC system is capable of almost completely suppressing the 50~Hz and its third harmonic, 150~Hz. This corresponds to 50~dB and 40~dB amplitude suppression respectively with respect to the noise reference. Moreover, we find that the ANC performance is better than the conventional approach using a PID controller. Fig.~\ref{fig:anc_results} (b) shows the ANC performance for all 3 axial directions. One can observe from the figure that ANC is very effective in cancelling the harmonic noise in the DC$-$1~kHz range with biggest noise reduction at lower frequencies. The noise performance is further broken down over different bandwidths, see Table~\ref{tab:2}. The 1~kHz bandwidth corresponds anti-aliasing terminated bandwidth of the analog acquisition which roughly corresponds to the bandwidth of Mag690 fluxgate magnetometer. The maximum RMS noise suppression is around 35~dB in the 1~kHz band. Commercial SERF OPMs have a typical bandwidth of 150~Hz and a dynamic range of $\pm 5$~nT and require an ambient magnetic field to be below $50$~nT \cite{quspin}. As seen in Table~\ref{tab:2}, we achieve $3-7$~nT RMS noise in a 150~Hz bandwidth. With just a modest improvement of the performance of our ANC system, it would therefore be possible to operate a SERF OPM in unshielded conditions. We also note that the laboratory where the ANC system was deployed is exceptionally noisy due to high current experiments running in close proximity. In typical lab environments, the RMS magnetic field noise is in $<100$~nT range \cite{mag_canc,mag_can2}.

\subsection{Performance limits and considerations}
The performance of an FxLMS-based ANC system is limited by a number of factors. First, the control and sensing hardware imposes limits on the secondary path impulse response as well as the bandwidth. If the group delays introduced by the hardware are large resulting in a long impulse response the FxLMS algorithm will not be able to compensate well for the noise. This can be especially problematic if aggressive (high-order) anti-aliasing filters are used which introduce long group delays. To circumvent this, a low order low pass filter can be used. This, however, comes at the expense of having to increase the sampling rate (oversampling). This introduces further limitations due to the fact that there is an upper time limit of the processing speed of the FxLMS algorithm. Another way to counter the long delay response is to increase the filter length. However, this increases the processing time and resources on the FPGA hardware and reduces the maximum bandwidth. Moreover, whilst increasing the filter length reduces the steady state error, it degrades the rate of convergence process to reach the steady state error \cite{anc3}. \\

\begin{table}[t!]
\begin{ruledtabular}
\begin{threeparttable}

\caption{RMS noise performance of PID and ANC systems benchmarked against the environmental noise. Bottom table expresses the RMS magnetic field noise converted to frequency using the gyromagnetic ratio of Cesium, $\gamma=350$~kHz/G. Cesium is a common atomic species used in OPMs. The frequency values on the axis row correspond to the bandwidths for which the noise performance is calculated for.}
\label{tab:2}
\setlength\tabcolsep{0pt} 

\begin{tabular*}{\columnwidth}{@{\extracolsep{\fill}} l cccccc}
\toprule
      & \multicolumn{2}{c}{Noise} & \multicolumn{2}{c}{PID}  & \multicolumn{2}{c}{ANC} \\ 
    
    \cmidrule(lr){2-3} \cmidrule(lr){4-5} \cmidrule(lr){6-7} 

    Axis & 1~kHz & 150~Hz & 1~kHz & 150~Hz & 1~kHz & 150~Hz \\ \colrule\\
    
\midrule
     x-axis & 800 nT & N/A & 112 nT & 80 nT & 16 nT & 7 nT \\
     y-axis & 150 nT & N/A & 35 nT & 25 nT & 7 nT & 3 nT \\
     z-axis & 450 nT & N/A & 23 nT & 16 nT & 12 nT & 7 nT \\
\bottomrule
\end{tabular*}
\vspace{0.5cm}

\begin{tabular*}{\columnwidth}{@{\extracolsep{\fill}} l cccccc}
\toprule
      & \multicolumn{2}{c}{Noise} & \multicolumn{2}{c}{PID}  & \multicolumn{2}{c}{ANC} \\ 
    
    \cmidrule(lr){2-3} \cmidrule(lr){4-5} \cmidrule(lr){6-7}

    Axis & 1~kHz & 150~Hz & 1~kHz & 150~Hz & 1~kHz & 150~Hz \\ \colrule\\
    
\midrule
     x-axis & 2800 Hz & N/A & 392 Hz & 280 Hz & 56 Hz & 24.5 Hz \\
     y-axis & 525 Hz & N/A & 123 Hz & 87.5 Hz & 24.5 Hz & 10.5 Hz \\
     z-axis & 1580 Hz & N/A & 80.5 Hz & 56 Hz & 42 Hz & 24.5 Hz \\
\bottomrule
\end{tabular*}
\end{threeparttable}
\end{ruledtabular}
\end{table}
Another highly crucial ingredient in determining the performance of the ANC system is the noise coherence. In real world environments multiple magnetic field noise sources may be present. These will have their own magnetic field noise profiles of varying amplitude, spectral profile and phase. Due to the principle of superposition and field decay, the resulting environmental noise will have spatial and temporal inhomogeneities. Moreover, if any conductive objects are present in the vicinity of the noise or error reference magnetometers, the time varying magnetic fields will induce eddy currents in the conductive objects which will produce their own respective magnetic fields \cite{mag_con}. As a result, the relative position of the noise and error reference magnetic field sensors can have a significant effect on the noise cancellation of the ANC system since it fundamentally relies on common-mode noise cancellation. The degree of noise coherence between the error and reference sensor can be quantified by the coherence function
\begin{equation}
    \gamma^2(f)=\frac{\left|S_{x,y}(f) \right|^2}{S_x(f)S_y(f)},
\end{equation}
where $S_x(f)$ and $S_y(f)$ are power spectral densities of the error and noise reference signals and $S_{x,y}(f)$ is the cross power spectral density. The coherence function takes the range
\begin{equation}
    0\leq \gamma^2(f) \leq 1,
\end{equation}
where a zero value of coherence implies no correlation between the signals and a value of 1 implies perfect signal correlation. Real-time estimation of the noise coherence between the noise reference and the error sensors can be used to determine the optimal placement of the noise reference magnetometer relative to the error magnetometer. The theoretical maximum cancellation of the noise for a given frequency band is given by \cite{morgan_kuo}
\begin{equation}
    \alpha(f)=-10\log_{10}\left(1-\gamma^2(f)\right)[\textrm{dB}].
\end{equation}
Active noise control systems applied to cancel the acoustic noise can inadvertently suffer with echo effects which occur as a result of the sound generated by the noise cancelling speaker propagating and reaching the noise reference microphone. Echo effects can also occur when attempting to cancel the magnetic field noise if the magnetic field produced by the coils is sensed by the noise reference magnetometer (which is located at a finite distance from the coils). If such fields are strong they can inadvertently introduce a feedback loop in the ANC system rendering it unstable. One solution is to place the noise reference sensor further away from the coils. However, this comes at the expense of reducing the noise coherence between the reference and error magnetometers \cite{mag_con}. An alternative is to introduce adaptive echo cancellation through primary path modelling which is then incorporated into the FxLMS algorithm \cite{adaptive_echo,anc_echo}. Another strategy to mitigate field echo contamination is to use coil geometries that have one-sided flux pattern which keeps the generated anti-noise field within the inner coil structure where the error sensor resides. This can be achieved through Halbach-type electromagnet architecture \cite{halbach,halbach_electromag}. Finally, the noise floor of the in-loop error and noise reference sensors will ultimately impose an upper limit on the noise suppression of the environmental field. \\ \\
The FxLMS algorithm used in this work is by far the most common approach deployed in active noise control. However, there exists an extensive family of alternative adaptive algorithms which possess a varying degree of capabilities ranging from the speed of convergence to the value of steady state error and computational complexity \cite{anc1,adaptive_algos}. Depending on the conditions for ANC, a different adaptive algorithm could be employed to enable better performance e.g. greater noise suppression at the expense of increased hardware and processing requirements. This also applies to secondary path modelling stage.

\section{Conclusions}
We have shown that active noise control is an effective technique for broadband magnetic field noise cancellation. It requires minimal hands on approach to tuning for optimal performance and is flexible for deployment across different hardware architectures. Moreover, in situations where the signals of interest are much smaller than the ambient noise, active noise control implementation can be used to increase the signal to noise ratio by increasing the resolution of the measurement via a reduced input range over the same number of bits. Finally, for highly sensitive devices such as optically pumped magnetometers where the intrinsic device sensitivity strongly depends on the ambient field noise, an active noise control system can be used to enable highly sensitive operation in magnetically unshielded environments.

\section*{Acknowledgements}
This work was supported by 
the UK Quantum Technology Hub in Sensing and Timing, funded by the Engineering and Physical Sciences Research Council (EPSRC) (EP/T001046/1), 
the Nottingham Impact Accelerator / EPSRC Impact Acceleration Account (IAA), 
and by Dstl via the Defence and Security Accelerator: www.gov.uk/dasa.
We thank Yonina Eldar for reading the manuscript.

\section*{Data availability}
The data that support the findings of this study are available from the corresponding author upon reasonable request.

\end{document}